\documentclass[aps,prb,a4paper,longbibliography,notitlepage,noshowpacs,10pt]{revtex4-1}
\usepackage{graphicx}
\usepackage{inputenc}
\usepackage{epstopdf}
\usepackage{siunitx}

\begin{document}

\title{Fundamental dissipation due to bound fermions in the zero-temperature limit}
\author{S.~Autti}
\email{s.autti@lancaster.ac.uk}
\affiliation{Department of Physics, Lancaster University, Lancaster LA1 4YB, UK.}
\author{R. P. Haley}
\affiliation{Department of Physics, Lancaster University, Lancaster LA1 4YB, UK.}
\author{A. Jennings}
\affiliation{Department of Physics, Lancaster University, Lancaster LA1 4YB, UK.}
\author{G. R. Pickett}
\affiliation{Department of Physics, Lancaster University, Lancaster LA1 4YB, UK.}
\author{R. Schanen}
\affiliation{Department of Physics, Lancaster University, Lancaster LA1 4YB, UK.}
\author{V. Tsepelin}
\affiliation{Department of Physics, Lancaster University, Lancaster LA1 4YB, UK.}
\author{J. Vonka}
\email{Current address: Paul Scherrer Institute, Forschungsstrasse 111, 5232 Villigen PSI, Switzerland}
\affiliation{Department of Physics, Lancaster University, Lancaster LA1 4YB, UK.}
\author{T. Wilcox}
\affiliation{Department of Physics, Lancaster University, Lancaster LA1 4YB, UK.}
\author{D. E. Zmeev}
\affiliation{Department of Physics, Lancaster University, Lancaster LA1 4YB, UK.}
\author{A. A. Soldatov}
\affiliation{P.L. Kapitza Institute for Physical Problems of RAS, 119334 Moscow, Russia}

%PRL length guidelines: main text max 3750 words including captions and figure size in text equivalent, abstract<600 characters
% Theorists potentially interested in giving comments/refereeing: Jim Sauls

%Nature communications: (1) abstract less than 150 words with references; (2) main text 5000 words or less; (3) sections Introduction, Resutls, Discussion, Methods [<3000 words]; (4) Figure legends [<350 words]

\begin{abstract}
The ground state of a fermionic condensate is well protected against perturbations in the presence of an isotropic gap. Regions of gap suppression, surfaces and vortex cores which host Andreev-bound states, seemingly lift that strict protection. Here we show that the role of bound states is more subtle: when a macroscopic object moves in superfluid $^3$He at velocities exceeding the Landau critical velocity, little to no bulk pair breaking takes place, while the damping observed originates from the bound states covering the moving object. We identify two separate timescales that govern the bound state dynamics, one of them much longer than theoretically anticipated, and show that the bound states do not interact with bulk excitations. 

\end{abstract}

\maketitle

\section*{Introduction}

Interfaces in fermionic condensates and the bound states hosted by them have recently attracted research focus in numerous condensed matter systems. Contemporary superfluid research is moving towards nanoscale probes\cite{PhysRevB.100.020506,PhysRevB.101.060503}, nanoconfinement\cite{PolarDmitriev,HQVs_prl,Levitin841}, and microscopic dissipation mechanisms related to quantum turbulence\cite{Silaev_PhysRevLett.108,Makinen_PhysRevB.97.014527}. Unconventional superconductivity in various materials\cite{Sauls2018} and the dynamics of certain atomic BECs\cite{Yefsah2013426} is characterised by the physics of bound excitations. Nanofabrication of superconducting hybrid structures has enabled tailoring bound states with desired properties. For instance, engineered superconducting systems containing Andreev-bound states (ABS) are used in studies of Majorana fermions\cite{Majorana1,Li2018875} aimed at producing components of, say, a quantum computer\cite{Mourik1003}. In this article we show that bound states covering the surface of a fermionic condensate not only determine the properties of its surfaces, but ABS dynamics are also central in understanding one of the most elementary of bulk phenomena: the drag an object experiences when moving at constant velocity through the fermionic superfluid.

A recent experiment showed, contrary to the classic textbook picture, that a macroscopic object moving quasi-uniformly in superfluid $^3$He-B can exceed the predicted Landau velocity \cite{PhysRev.60.356} in the zero-temperature limit without destroying the condensate\cite{Bradley2016}. Phenomenologically the dissipation experienced by objects moving rapidly can be explained in terms of the Lambert picture\cite{LAMBERT1992294}. This scenario is based on assuming that there is a well-defined 2D gapless spectrum of bound states available within roughly a coherence length of all surfaces, and that the fermions occupying these states can escape to the bulk promoted by macroscopic flow fields, thereby providing dissipation. Simultaneously the surface layer shields the bulk condensate from interacting with the moving object, preventing pair breaking even if the object's velocity exceeds the Landau velocity. Assuming a free 2D gas of surface excitations is clearly a simplification, but the model captures the main features of the experiment well enough for the presentation of the results.

The above approach still lacks rigorous theoretical justification. In particular, its critics have pointed out that thermalisation of the surface-bound states should be so fast, of the order of nanoseconds, that the equilibrium cannot be disturbed by the relatively slow motion (typically milliseconds) of a macroscopic moving object\cite{thuneberg2018}. Here we provide experimental evidence to end the controversy. We show that this criticism correctly identifies only one of the two timescales that govern the dynamics of surface-bound states, and that this process is very fast as theoretically anticipated. Crucially, we also observe a second, much slower process, which allows for the surface states' contribution to dissipation. Our experiment shows, remarkably, that the observed dissipation is independent of the density of thermal bulk excitations. This observation justifies the assumption that surface-bound states are involved in the process in the first place.

An ideal experiment aimed at probing ABS-originating dissipation in superfluid $^3$He-B would be carried out at a temperature so low that thermal bulk excitations vanish completely. Otherwise trivial collisions with them provide a thermal dissipation background. In our experiments  the sample temperature was between \SI{140}{\micro\kelvin} and \SI{240}{\micro\kelvin}. Therefore the density of thermal bulk quasiparticles is not negligible, but it is low enough that they propagate ballistically, and their contribution to observed dissipation can be confidently subtracted. 

In these conditions at saturated vapour pressure the Landau critical velocity is $v_\mathrm{L}=\Delta/p_\mathrm{F}=27$~mm/s ($\Delta$ is the superfluid gap and $p_\mathrm{F}$ the Fermi momentum). We can reach and exceed this velocity in both AC and DC measurements by driving a large current through a goalpost-shaped superconducting wire (legs 25~mm, crossbar 9~mm, thickness \SI{135}{\micro\meter}). An external magnetic field is oriented parallel to the legs of the wire, and the resulting Lorentz force moves the wire. The wire is surrounded by a volume of superfluid $^3$He-B, and the drag force experienced by the wire is inferred from the number of emitted quasiparticles, seen as a heat pulse measured by the thermometers. 

\section*{Results}

Let us study the consequences of a gapless spectrum of bound states covering the surface of the moving wire crossbar at a temperature so low that there are no thermal bulk excitations. In the presence of macroscopic flow, both bulk and gapless surface dispersion curves for quasiparticles (and holes) move according to $\pm v_{\rm fl} p_\mathrm{F}$ (in one dimension). Here $v_{\rm fl}$ is the local flow velocity of the superfluid. In the frame of the crossbar, the bulk far away from the wire flows at $v_{\rm fl}=-v$, where $v$ is the velocity of the crossbar in the laboratory frame. Near the cylindrical crossbar the flow is enhanced and reaches a maximum of $v_{\rm fl}=-2v$, assuming perfect potential flow. These two contributions narrow down the energy mismatch $\Delta$ between available bulk states and populated ABS states by $3v p_\mathrm{F}$, and the flow therefore enables ABS escape from the surface of the wire to the bulk when the wire moves faster than $v_\mathrm{c}=1/3 \cdot \Delta/p_\mathrm{F} =9$~mm/s, the well known critical velocity in AC measurements\cite{AC_Landau}. Strikingly, if $v$ is held constant, the ABS spectrum finds an equilibrium and the drag force disappears\cite{Bradley2016}.

Such difference between oscillatory and uniform motion must arise from the time constants that govern the process described above. Basically, the repeating reversal of direction in oscillatory motion replenishes the reserve of quasiparticles that can escape to bulk when the velocity exceeds the critical velocity, while in uniform motion such replenishment is not available. If we compare equally-long and otherwise identical trajectories where velocity is either reversed or not, the reversal should increase the observed dissipation. In the absence of the bound state contribution such directionality dependence is not expected.\cite{Bradley2016}

\begin{figure*}[tb!]
	\centerline{\includegraphics[width=\linewidth]{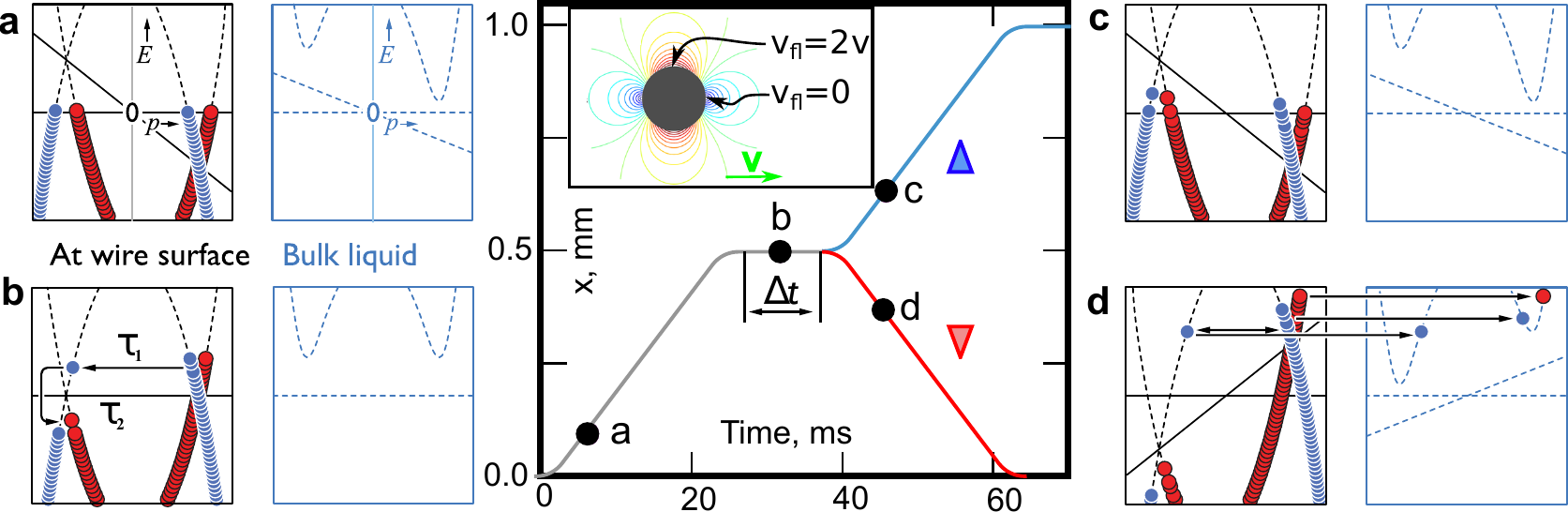}}
	\caption{{\bf Schematic illustration of the measurement at zero temperature:} (Centre) The crossbar of the wire is moved in two phases. Comparing two different, equally long trajectories reveals the bound state dissipation. Bound state dynamics is probed by varying the waiting time $\Delta t$ between the two phases. All ramps begin by accelerating the wire from zero to a constant velocity (here $v=40~$mm/s), which in the coordinate system of the wire creates a flow around the wire. The flow pattern is shown with contour lines in the inset. The flow shifts both the bound state spectrum and the available bulk states as described in the main text. After the acceleration, the population of bound quasiparticles (red circles) and quasiholes (blue circles) on the wire surface reaches a steady state (panel {\bf a}). The dispersion curves are drawn for the top (or bottom) generatrices of the wire, where the flow velocity is maximal. During $\Delta t$, momentum exchange with the wire surface allows exchange of bound quasiparticle populations between the branches ($\tau_1$). Within a branch the population relaxes with $\tau_2$ (down ramp, panel {\bf b}). If the first phase of motion is followed by the second one in the same direction and $\Delta t \lesssim \tau_1$ (up ramp, panel {\bf c}), the remaining imbalance in populations results in less dissipation from quasiparticles escaping to bulk than a fully symmetric one would. If the direction of motion is reversed (panel {\bf d}) and $\Delta t \lesssim \tau_1$, then the dissipation will be enhanced. }
	\label{ramps_schematic} 
\end{figure*}

In order to experimentally substantiate this picture, and to remove the effect of the finite density of thermal bulk excitations that cannot be avoided in experiments, we move the wire crossbar in two phases (Fig.~\ref{ramps_schematic}). Each measurement starts from standstill long enough to remove any history dependence. The wire is then quickly accelerated to $v$, the velocity is kept constant for 0.5~mm distance, before reducing the velocity back to zero\cite{Zmeev2014}. The acceleration time was varied from 3~ms to 5~ms, and the obtained results were independent of it. After time $\Delta t$, the ramp is repeated from this new starting location with either $v$ (``up ramp'') or $-v$ (``down ramp''). This allows subtracting the dissipation measured for up ramps $Q_\mathrm{up}$ from that measured for down ramps $Q_\mathrm{down}$, $\Delta Q=Q_\mathrm{down}-Q_\mathrm{up}$. As described in detail in Methods, the dissipation due to ambient bulk quasiparticles is cancelled assuming the up and down trajectories are equally long. 

The above measurement shows that bound-state dissipation is characterised by two regimes: First, the difference in the measured dissipation, $\Delta Q$, is zero when $v\leq9$~mm/s (Fig.~\ref{velocity_dep}b in Methods). This implies that, in the zero-temperature limit, dissipation vanishes below the critical velocity $v_\mathrm{c}=9~$mm/s. Second, for velocities higher than 9~mm/s at $\Delta t=0$, up ramps experience less dissipation than down ramps. Summing over the solid angle of allowed escape directions to bulk and over the available bound quasiparticle states on the wire surface (see Methods) yields the power law $\Delta Q \propto ((v-v_\mathrm{c})/v_\mathrm{c})^{2.5}$, which is in good agreement with the experiment. We emphasise that $v_\mathrm{c}=9$~mm/s is the well known AC critical velocity, confirming that the results of the DC measurement directly apply to dissipation experienced by moving objects in superfluid $^3$He-B in general. This confirms that direct bulk pair breaking is replaced by a much weaker drag force originating from surface-bound quasiparticles, as speculated earlier\cite{Bradley2016}.

\begin{figure}[tb!]
	\includegraphics[width=\linewidth]{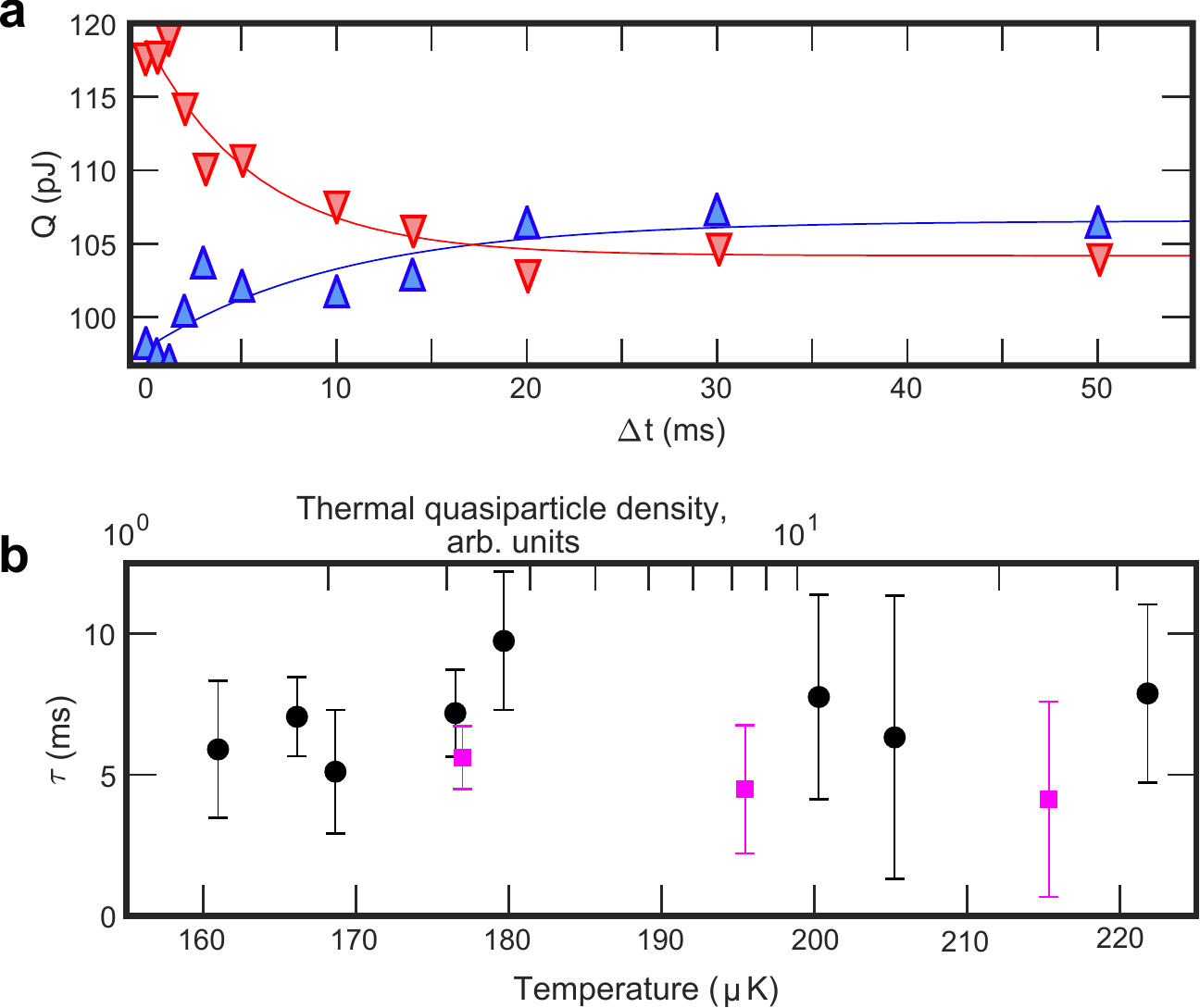}
	\caption{{\bf Bound state dynamics:} ({\bf a}) Measured dissipation as a function of $\Delta t$ for up ramps (blue upward triangles) and down ramps (red downward triangles) reveals the characteristic time of bound-state dynamics on the wire surface. The solid lines are exponential fits to the data. At large waiting times $\Delta Q \approx -2.6$~pJ, as expected due to the hysteretic contribution analysed in Fig.\ref{velocity_dep}. ({\bf b}) The fitted time constants with $^4$He preplating (magenta squares) and without it (black circles) are independent of temperature. The relative change in thermal quasiparticle density is shown on the top axis. All data in this figure was measured at $H=130$~mT and $v=45~$mm/s.}
	\label{wait_dep} 
\end{figure}

We can now vary $\Delta t$ to measure the dynamics of bound states on the wire surface (Fig.~\ref{wait_dep}). The difference in dissipation, $\Delta Q$, disappears exponentially as $\exp(-\Delta t / \tau)$, where $\tau\approx 6~$ms in the measured temperature range of \SI{160}{\micro\kelvin}~--~\SI{230}{\micro\kelvin}. Within this range of temperatures the thermal bulk quasiparticle density varies by almost two orders of magnitude, while measured $\tau$ is constant. Together with the above observations this justifies the assumption that bound quasiparticles are responsible for the dissipation. It also rules out any direct interaction between surface-bound and bulk quasiparticles as the source of thermalisation. 

The dynamics of bound quasiparticles on the wire surface can be described by two time constants in our toy model. The first one, $\tau_1$, gives the probability of the inter-branch process where a quasiparticle scatters from the wire exchanging momentum going from, say, $p$ to $-p$. This process enables drag, as the exit channel to bulk is open only in the direction of wire motion while only quasiparticles with the momentum in the opposite direction gain energy due to the flow. That is, drag is produced by those quasiparticles that escape to bulk right after scattering with the wire, removing momentum from the wire. The second time constant, $\tau_2$, describes the rate at which quasiparticles that did not escape to bulk relax within each dispersion branch to the thermal equilibrium distribution distorted by process one. If $\tau_2=\infty$, then the imbalance produced by moving the wire and stopping in the middle, measured by comparing up and down ramps, disappears diffusively as determined by $\tau_1$. Assuming $\tau_2=0$, the imbalance still disappears at the rate given by $\tau_1$, but this time the process is deterministic. Therefore, regardless of $\tau_2$, we have measured $\tau_1\approx\tau=6~$ms. 

\begin{figure}[htb!]
	\includegraphics[width=\linewidth]{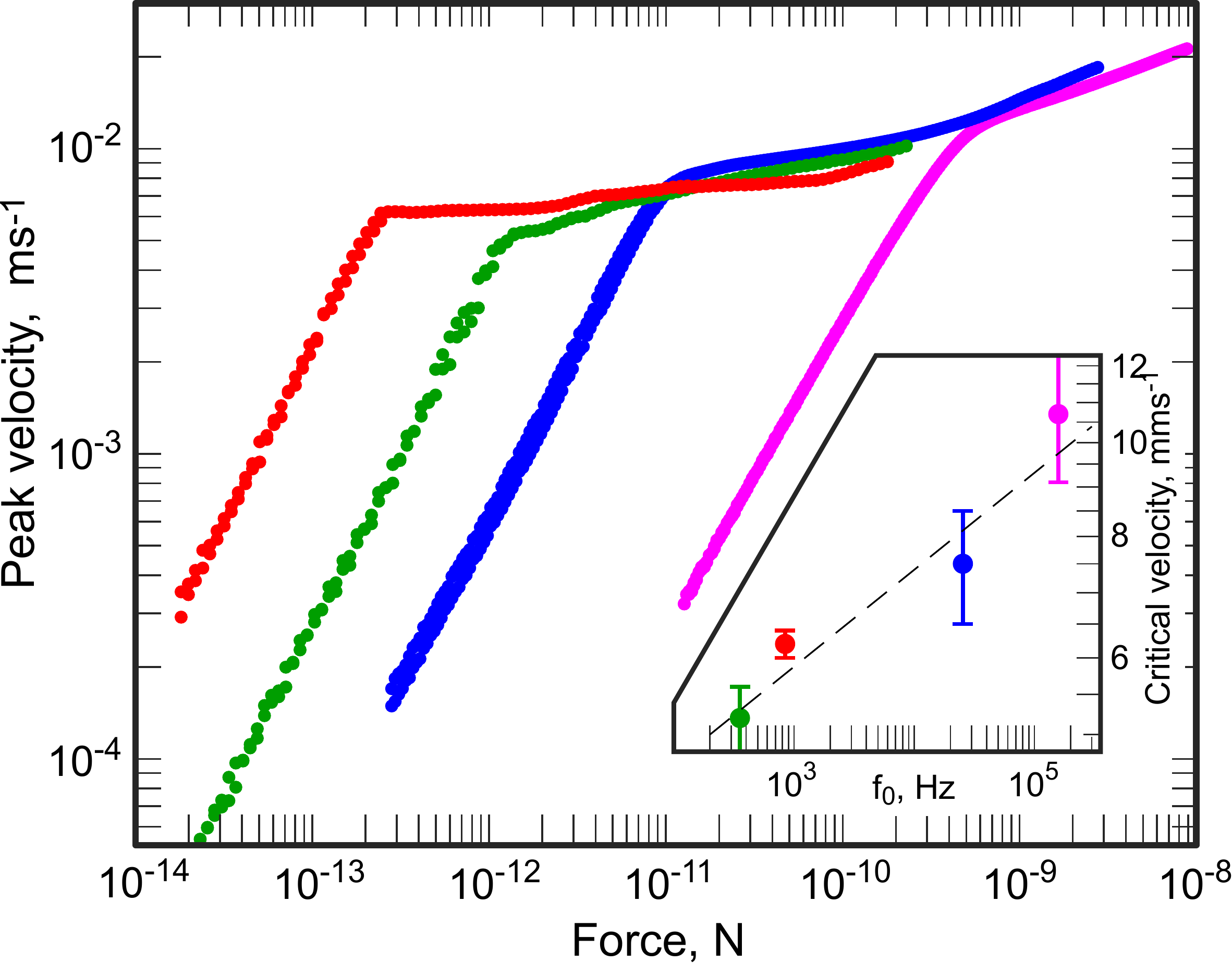}
	\caption{{\bf AC critical velocity measurements as a function of frequency:} In AC measurements, critical velocity $v_\mathrm{c}$ is seen as a sharp increase in the force needed to increase peak oscillation velocity above the critical velocity. The exact critical velocity depends on details of the flow field around the probe, and therefore on the shape of the object. The measured probes are, in order from left to right: a \SI{1}{\micro\meter} thick vibrating wire operating at 843~Hz (red points, measured at \SI{110}{\micro\kelvin}), a \SI{4.5}{\micro\meter} thick vibrating wire at 355~Hz (green points, \SI{110}{\micro\kelvin}), a custom-made quartz tuning fork at 25.7~kHz (blue points, \SI{120}{\micro\kelvin}), and an overtone of the same fork at 158kHz (magenta points,\SI{110}{\micro\kelvin}). The inset shows critical velocities extracted from the main figure. Black dash line is a guide to the eye that corresponds to $v_\mathrm{c}\propto f_0^{0.1}$. }
	\label{freq_dep} 
\end{figure}

It is also possible to set an experimental upper bound for $\tau_2$. We have measured the AC critical velocity of a set of probes with resonance frequencies $f_0$ ranging from 350~Hz to 158~kHz (Fig.~\ref{freq_dep}). The critical velocity is seen as kink in the velocity of the probe measured as a function of force. It corresponds to a sudden increase in the force required to increase the velocity by a given amount. As speculated by Lambert\cite{LAMBERT1992294}, when $f_0 > 1/\tau_2$, one should see a clear reduction in the observed AC critical velocity as quasiparticles would be able to ``climb up'' the dispersion curves diffusively by scattering back and forth between $p$ and $-p$. Say, by a two-step process one would get escape to bulk starting at  4.5~mm/s, and so on. 

We observe no critical velocity reduction up to $f_0=158~$kHz. In fact the observed critical velocity slowly and smoothly increases as a function of resonance frequency. This happens despite the varying geometries of the four probes used: the two low frequency probes are vibrating superconducting wires, while the two kHz probes are the first harmonic and an overtone of a custom-made quartz tuning fork with smooth surfaces. While studying the reason for the observed slow power-law increase of $v_\mathrm{c}$  ($v_\mathrm{c} \propto f_0^{0.1}$) systematically remains a task for the future, this result implies that $\tau_2\lesssim 1/f_0 \approx$\SI{10}{\micro\second}. We believe that $\tau_2$ describes the fast thermalisation process anticipated in Ref.~\citenum{thuneberg2018}.

Surface specularity may play an important role in the bound state escape process. In the B phase the gap suppression does not significantly depend on the details of surface scattering, and therefore the bound states' spectrum itself is approximately independent of specularity\cite{Nagato19981135,Nagai2008}. On the other hand, one would expect that $\tau_1$ depends on specularity as it describes the process of quasiparticle scattering with the wire. It is known that the surface scattering can be tuned from diffuse to specular continuously by preplating the sample with slightly over two monolayers of $^4$He\,\,\cite{PhysRevB.47.319}. It is possible that $^4$He coverage also changes the scattering qualitatively by removing magnetic spin exchange with the surface layer\cite{Fomin2018}. However, reaching fully specular scattering requires a 2D superfluid layer of $^4$He, which flows to the coldest part of the sample container (the heat sink) and cannot therefore be stabilised on the wire in our experiment\cite{Levitin841}. We have resorted to measuring the effect of adding two monolayers of solid $^4$He on the wire surface (and all other surfaces in the sample container). While this causes the heat exchangers to become less efficient\cite{Sinters2020}, we observe very little change in the bound state dynamics or dissipation. In particular, $\tau_1$ remains $\approx5$~ms within the scatter of the data. This shows that the quasiparticle escape process is robust against at least small changes in the scattering conditions. 

\section*{Discussion}

In conclusion, our experiments confirm that quasiparticles (and holes) bound to the region of gap suppression near solid surfaces in the B phase of superfluid $^3$He are responsible for the zero-temperature dissipation experienced by macroscopic objects moving in the superfluid. The dissipation begins when the flow velocity is sufficient for releasing the bound quasiparticles to the bulk of the superfluid. We have demonstrated experimentally that the bound states re-equilibrate through a two-stage process involving a fast thermalisation step with $\tau_2<$\SI{10}{\micro\second}, but importantly also a theoretically unforeseen slow scattering step characterised by $\tau_1=6~ $ms. To be clear, $\tau_1$ is the effective time constant describing the process where the quasiparticle momentum is changed from $\mathbf{p}_\mathrm{F}$ to $-\mathbf{p}_\mathrm{F}$. In a proper two-dimensional treatment of the process that $\tau_1$ emerges from, this may involve a large number of scattering events which collectively allow the quasiparticle to eventually find the very opposite momentum. We speculate that this process is potentially the underlying reason as to why $\tau_1$ is not of the order of nanoseconds. Furthermore, we vary the bulk quasiparticle concentration to show that the bulk states do not contribute to bound state dynamics on any observable timescale. Therefore our results can be readily generalised to the description of transient phenomena in 2-dimensional Dirac systems such as graphene\cite{PhysRevB.94.125445}, if the escape process to bulk is neglected. It is worth emphasising that the bulk escape process is temperature independent at low temperatures and, hence, provides dissipation even in the zero-temperature limit.

We acknowledge the preliminary nature of the bound-state model presented above. In particular, it remains an open question how exactly bound state dynamics should be described in a full 2-dimensional model of the wire surface with a flow velocity distribution, diffusion or transport between various parts of the surface etc. Competing theoretical suggestions are sparse, but there have been some discussions related to a layer of vortices covering the wire acting as a buffer and therefore masking the Landau velocity\cite{thuneberg2018}. On the other hand, theoretical work on superflows exceeding the Landau velocity have recently been published studying other systems, such as polaron-polaritons\cite{nielsen2020superfluid} and graphene\cite{PhysRevLett.121.136804}. For these reasons it remains of interest to provide additional experimental insight. For example, one could change the Landau velocity by studying flow in a confined geometry, say, in the recently-discovered polar phase of superfluid $^3$He \cite{PolarDmitriev}. The gap spectrum of the polar phase contains a nodal line, meaning that the Landau speed limit is zero in that plane. On the other hand, it seems essential to enhance the experiment presented in this article by building a detector or spectrometer fast enough to distinguish the dissipation associated with the separate phases of the wire motion. This requires designing and devising nano-sized instruments \cite{PhysRevB.100.020506,PhysRevB.101.060503}.

\section*{Author contributions}

All authors contributed to gathering the results and writing the manuscript.

\section*{Acknowledgements}
 We thank Erkki Thuneberg for stimulating discussions. This work was funded by UK EPSRC (grant No.EP/P024203/1) and EU H2020 European Microkelvin Platform (Grant Agreement 824109). We acknowledge M.G. Ward and A. Stokes for their excellent technical support. S.A. acknowledges financial support from the Jenny and Antti Wihuri Foundation.
 
\section*{Competing interests}
 The authors declare no competing interests
 
\section*{Methods}
 
An ideal experiment aimed at probing bound-state-originating dissipation in superfluid $^3$He-B would be carried out at a temperature so low that thermal bulk excitations are rare enough to be neglected completely. Otherwise, trivial collisions with them provide a thermal dissipation background. In our experiments we used a nested nuclear demagnetisation cryostat \cite{PICKETT200375}, capable of holding the sample temperature between \SI{100}{\micro\kelvin} and \SI{200}{\micro\kelvin} for several days. Temperature was measured with a quartz tuning fork \cite{2007_forks}, and a vibrating wire thermometer \cite{PhysRevB.57.14381,Guenault1986}. Pressure in all measurements was the saturated vapour pressure, corresponding to superfluid transition temperature $T_\mathrm{c}=$\SI{929}{\micro\kelvin}. Therefore the density of thermal bulk quasiparticles is not negligible, but it is low enough so that they propagate ballistically, and their contribution to the observed dissipation can be subtracted. We can vary the thermal quasiparticle density by two orders of magnitude within the ballistic regime. Subtracting the thermal quasiparticle contribution is robust and reliable because the gap spectrum of the B phase is isotropic in zero magnetic field, and only slightly distorted in small external magnetic field.

The main probe used in the experiment is the goalpost-shaped wire. The bolometric volume that surrounds the wire is calibrated by resonant AC measurements\cite{MSkyba_phd} by fitting the measured calibration data to known BCS heat capacity using the effective volume of the sample as a fitting parameter. The fitted volume is $16~\mathrm{cm}^3$, which falls between the free volume of the sample container, $15~\mathrm{cm}^3$, and the total volume of the sample container including the volume within the heat exchangers, $32~\mathrm{cm}^3$. 

We simultaneously monitor the position of the wire by picking up a high-frequency signal mixed in with the driving current using nearby coils. This method is not sensitive enough for measuring the drag force\cite{Bradley2011114}, but together with the known wall-to-wall distance ($\approx10~$mm) it calibrates the range of motion. Finally, we record the induced voltage across the wire with a 4-point measurement. The induced voltage reveals whether the main AC resonance or some higher mode of oscillation is excited by the DC drive, allowing us to ensure that dissipation due to these modes was minimised during all measurements.
 
We access zero-temperature dissipation by moving the wire crossbar in two phases (Fig.~\ref{ramps_schematic}). Each measurement starts from standstill long enough to remove any history dependence. The wire is then quickly accelerated to $v$, the velocity is kept constant for 0.5~mm distance, before decelerating back to zero. After time $\Delta t$, the ramp is repeated from this new starting location with either $v$ (``up ramp'') or $-v$ (``down ramp''). This allows subtracting the dissipation measured for up ramps $Q_\mathrm{up}$ from that measured for down ramps $Q_\mathrm{down}$, $\Delta Q=Q_\mathrm{down}-Q_\mathrm{up}$.
 
\begin{figure}[tb!]
	\includegraphics[width=\linewidth]{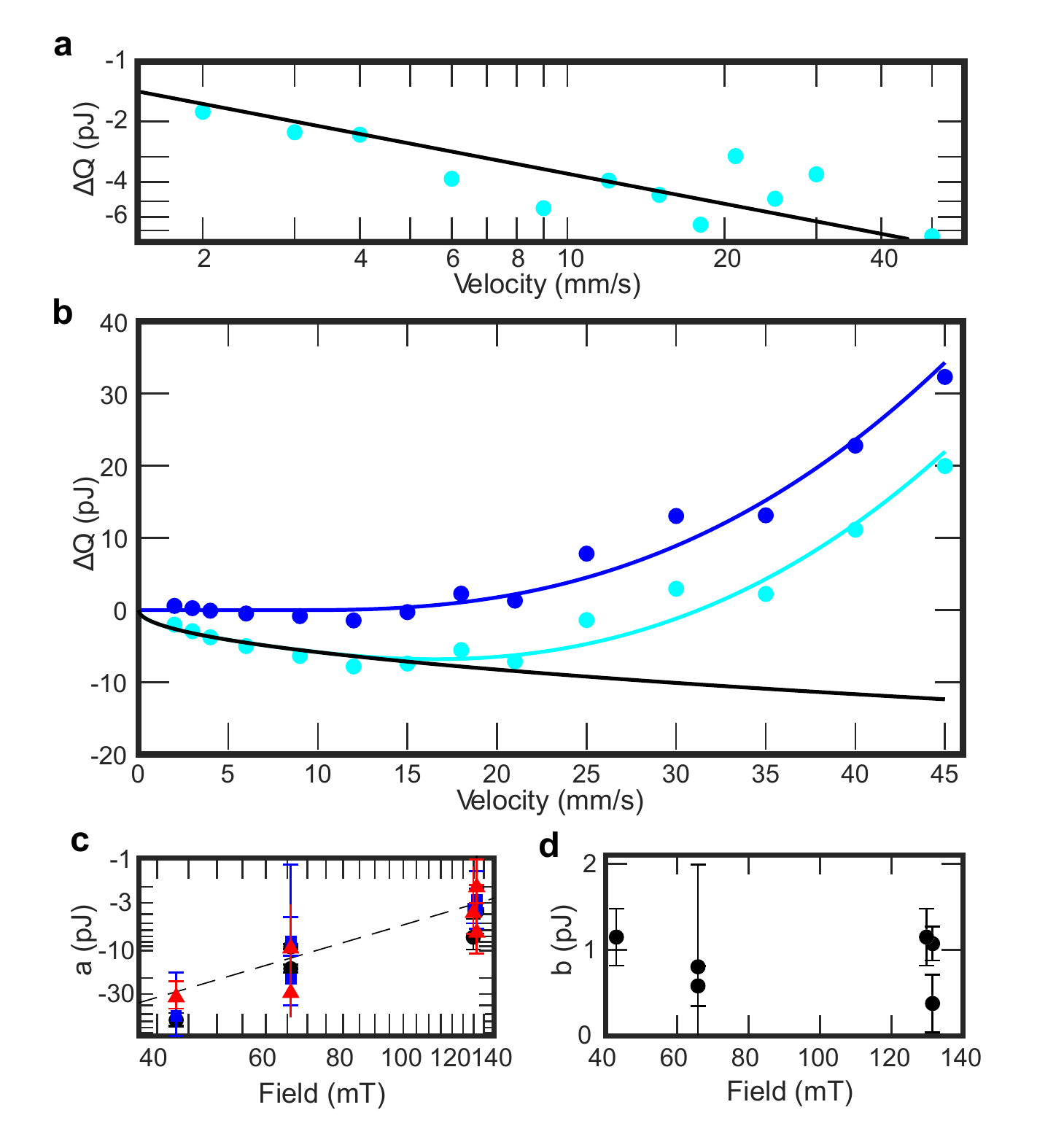}
	\caption{{\bf Bound state dissipation and critical velocity:} ({\bf a}) The measured hysteretic difference in dissipation with $\Delta t=50$~ms (cyan circles) follows the empirical power law $\Delta Q = a (v/v_\mathrm{c})^{1/2}$  (solid line), as clearly seen in log scale. ({\bf b}) When $\Delta t=0$ (cyan circles), the hysteretic contribution can be removed by fitting the data at $v\leq9$~mm/s with the above power law (solid black line). After subtracting this fit, the data (blue circles) follows $\Delta Q = b  ((v-v_\mathrm{c})/v_\mathrm{c})^{2.5}$ (solid blue line) above $v_\mathrm{c}=9~$mm/s, and $\Delta Q =0$ for $v<v_\mathrm{c}$, as explained in the text. The sum of the two fits is shown by the cyan line. The fitted values of $a$ are shown in panel ({\bf c}): Fits to data where $\Delta t=0$ are shown with black circles, $\Delta t=50$~ms corresponds to blue squares, and $\Delta t=100$~ms is shown as red triangles. The dashed line is a guide to the eye that corresponds to $a=-5.0 / H^2~\mathrm{nJ \,mT^2}$. Panel ({\bf d}) shows fitted values of $b$ (black circles), which within the scatter of the data is independent of $H$ as expected. The temperature varied from \SI{150}{\micro\kelvin} to \SI{190}{\micro\kelvin} in these measurements.}
	\label{velocity_dep} 
\end{figure}

Ideally, both the trajectories would meet identical dissipation from scattering of thermal quasiparticles in the bulk. Subtracting the dissipation measured for up ramps $Q_\mathrm{up}$ from that measured for down ramps $Q_\mathrm{down}$ would leave only the ABS contribution $\Delta Q=Q_\mathrm{down}-Q_\mathrm{up}$, which for long enough $\Delta t$ is expected to be zero. In practice, the motion of the wire is hysteretic due to the motion of flux lines in the superconductor, driven by the current used for moving the wire in the magnetic field\cite{Zmeev2014}. Therefore the crossbar travels a shorter distance with down ramps than with up ramps, and as a result scatters a different amount of thermal quasiparticles, corresponding to $\Delta Q<0$. The hysteretic contribution follows $\Delta Q = a (v/v_\mathrm{c})^{1/2}$ (Fig.~\ref{velocity_dep}a), where $a \propto(\Delta I)^2$ (Fig.~\ref{velocity_dep}c). That is, the force moving the wire is kept constant in all the measurements (for a given velocity $v$), carried out at different magnetic fields $H$. Hence, the change in the current passed through the wire $\Delta I \propto  1/H$.

Once the hysteretic contribution is subtracted, $\Delta Q=0$ when $v\leq9$~mm/s (Fig.~\ref{velocity_dep}b), corresponding to zero dissipation from the bound states. For velocities higher than 9~mm/s at $\Delta t=0$, up ramps experience less dissipation than down ramps. Assuming perfect potential flow and constant density of states around Fermi energy, one can integrate the total solid angle where the escape condition to bulk is fulfilled: The number of available quasiparticle states at a fixed location on the wire surface increases as proportional to $v-v_\mathrm{c}$. The solid angle of directions where escape to bulk is allowed from that given location also scales like $v-v_\mathrm{c}$. Finally, the surface area on the wire crossbar which contributes to the dissipation expands like $\sqrt{v-v_\mathrm{c}}$. Assuming the damping scales proportionally to the available solid angle for escape, this yields the power law $\Delta Q = b  ((v-v_\mathrm{c})/v_\mathrm{c})^{2.5}$. This is in good agreement with the measurements at $\Delta t=0$ with $b\approx 1.0 ~\mathrm{pJ}$ (Fig.~\ref{velocity_dep}d). 
 
\section*{Data availability}
All the data in this paper is available in Ref.~\citenum{data_container} including descriptions of the data sets.

\bibliography{Flopper_bibliography}

\newpage
%%%%%%%%%%%%%%%%%%%%%%%%%%%%%%%%%%%%%%%%%
%%%%%%%%%%%%%%%%%%%%%%%%%%%%%%%%%%%%%%%%%
%%%%%%%%%%%%%%%%%%%%%%%%%%%%%%%%%%%%%%%%%
%%%%%%%%%%%%%%%%%%%%%%%%%%%%%%%%%%%%%%%%%

%%%%%%%%%%%%%%%%%%%%%%%%%%%%%%%%%%%%%%%%%
%%%%%%%%%%%%%%%%%%%%%%%%%%%%%%%%%%%%%%%%%
%%%%%%%%%%%%%%%%%%%%%%%%%%%%%%%%%%%%%%%%%
%%%%%%%%%%%%%%%%%%%%%%%%%%%%%%%%%%%%%%%%%

%%%%%%%%%%%%%%%%%%%%%%%%%%%%%%%%%%%%%%%%%
%%%%%%%%%%%%%%%%%%%%%%%%%%%%%%%%%%%%%%%%%
%%%%%%%%%%%%%%%%%%%%%%%%%%%%%%%%%%%%%%%%%
%%%%%%%%%%%%%%%%%%%%%%%%%%%%%%%%%%%%%%%%%

%%%%%%%%%%%%%%%%%%%%%%%%%%%%%%%%%%%%%%%%%
%%%%%%%%%%%%%%%%%%%%%%%%%%%%%%%%%%%%%%%%%
%%%%%%%%%%%%%%%%%%%%%%%%%%%%%%%%%%%%%%%%%
%%%%%%%%%%%%%%%%%%%%%%%%%%%%%%%%%%%%%%%%%
\end{document}